\documentclass[conference]{IEEEtran}

\usepackage[table]{xcolor}
\usepackage{times}
\usepackage{epsfig}
\usepackage{amsmath}
\usepackage{amsfonts}
\usepackage{graphicx}
\usepackage{amssymb}
\usepackage{amstext}
\usepackage{latexsym}
\usepackage{color}
\usepackage{ifthen}
\usepackage{multirow}
\usepackage{verbatim}
\usepackage{array,tabularx}
\usepackage{arydshln}
\usepackage[mathscr]{euscript}
\usepackage{accents}
 \usepackage{cite}
\usepackage{hhline}
\usepackage{caption}
\usepackage{subcaption}
\usepackage{enumerate}
\usepackage{xcolor}
\usepackage{mathtools}
\usepackage{url}
\usepackage{xparse}
\usepackage{makecell}
\usepackage{varwidth}

\newcolumntype{C}[1]{>{\centering\let\newline\\\arraybackslash\hspace{0pt}}m{#1}}

\usepackage[normalem]{ulem}

\usepackage{amsmath,pgfplots,amssymb}
\usepackage{subcaption,graphicx}

\newcommand{\be}{\begin{equation}}
\newcommand{\ee}{\end{equation}}
\newcommand{\bear}{\begin{eqnarray}}
\newcommand{\eear}{\end{eqnarray}}
\newcommand{\bears}{\begin{eqnarray*}}
\newcommand{\eears}{\end{eqnarray*}}
\newcommand{\bi}{\begin{itemize}}
\newcommand{\ei}{\end{itemize}}
\newcommand{\ben}{\begin{enumerate}}
\newcommand{\een}{\end{enumerate}}

\usepackage{tikz}
\usetikzlibrary{shapes.geometric}
\usetikzlibrary{shapes,snakes,patterns}
\usetikzlibrary{arrows,positioning,automata,calc}
\usetikzlibrary{plotmarks}

\definecolor{ogreen}{rgb}{0,0.5,0}
\definecolor{magenta}{rgb}{1.0, 0.11, 0.81}
\definecolor{mulberry}{rgb}{0.77, 0.29, 0.55}
\definecolor{xgray}{rgb}{0.5, 0.5, 0.5}
\definecolor{ao}{rgb}{0.0, 0.5, 0.0}
\definecolor{amber}{rgb}{1.0, 0.75, 0.0}
\definecolor{capri}{rgb}{0.0, 0.75, 1.0}
\definecolor{chocolate}{rgb}{0.91, 0.41, 0.17}

\newtheorem{theorem}{Theorem}

\newtheorem{property}{Property}

\newtheorem{example}{Example}
\newtheorem{definition}{Definition}

\newcommand*{\rowstyle}[1]{
  \gdef\@rowstyle{#1}%
  \leavevmode\@rowstyle
  \ignorespaces
}

\newcolumntype{=}{
  >{\gdef\@rowstyle{}\ignorespaces}%
}
\newcolumntype{+}{
  >{\leavevmode\@rowstyle\ignorespaces}%
}

\newenvironment{contexample}{
   \addtocounter{example}{-1} \begin{example}[continued]}{
   \end{example}}

\IEEEoverridecommandlockouts

\begin{document}
\title{Robust Private Information Retrieval on Coded Data}
\author{Razane Tajeddine, Salim El Rouayheb\\ ECE Department, IIT, Chicago\\ Emails: rtajeddi@hawk.iit.edu, salim@iit.edu \vspace{-1.5em} \thanks{This work was supported in part by NSF Grant CCF 1652867.}}

\maketitle 

\begin{abstract}
We consider the problem of designing PIR scheme on coded data when certain nodes are unresponsive. We provide the construction of $\nu$-robust PIR schemes that can tolerate up to $\nu$ unresponsive nodes. These schemes are adaptive and universally optimal in the sense of achieving (asymptotically) optimal download cost for any number of unresponsive nodes up to $\nu$.
\end{abstract}

\section{Introduction}

Consider a user who wishes to download a certain file from a distributed storage system (DSS) while keeping the identity of this file private. The user's concern about his/her privacy is due to many causes, such as concern about  surveillance, protection against online profiling from companies, etc. Private information retrieval (PIR) schemes \cite{PIR1995, chor1998private} allow a user to achieve privacy by querying the different nodes in the system, while guaranteeing that no information is being revealed about which   file is being retrieved.
A straightforward PIR scheme  consists of the user downloading all the files in the DSS, achieving perfect privacy. However, it has a very high communication cost. The literature on PIR has focused on efficient schemes that  can achieve privacy while  minimizing  different system costs, and in particular, the communication cost, which has received the most attention \cite{gasarch2004survey}.

Since its introduction in \cite{PIR1995}, the model of PIR assumes the data to be replicated on multiple nodes (e.g. \cite{sun2016capacitynoncol, sun2016capacity, yekhanin2010private, beimel2001information, beimel2002breaking}). 
Recently, there has been a growing interest in using codes in DSS to minimize the storage overhead of data. This has motivated recent works on PIR schemes for data stored under coded form and not just replicated \cite{shah2014one, chan2014private, tajeddine2016private, extended, banawan2016capacity, fazeli2015pir, blackburn2016pir, freij2016private}. The next example, taken from \cite{tajeddine2016private, extended},  illustrates the construction of a  PIR scheme on coded data.

\begin{example} \label{sec:introex}
Consider a DSS with $n=4$ nodes storing $m$ files $X_i = (a_i,b_i),$ $a_i,b_i\in GF(3^\ell),$ $i=1,2,\dots, m$. The files are stored using an $(n,k)=(4,2)$ MDS code over $GF(3)$. Let $A = 
\left[\begin{array}{cccc}
a_1 & a_2 & \dots & a_m
\end{array} \right]^T$, and $B = 
\left[\begin{array}{cccc}
b_1 & b_2 & \dots & b_m
\end{array} \right]^T$ represent the first and second half (block) all the files in the DSS, respectively. Nodes $1,\dots, 4$ store $A, B, A+B, A+2B$, respectively. The user requires to retrieve file $X_f = (a_f,b_f)$, $f \in \{1,\dots, m\}$, privately, by querying the four nodes, but without revealing any information about the file index $f$ to any of them.  In the scheme in \cite{extended, tajeddine2016private}, the user generates an iid random vector $\mathbf{u}=\left[\begin{array}{ccc}
u_1 & \dots &u_m
\end{array}\right]^T$ with elements chosen uniformly at random from $GF(3)$ and independent of $f$, and forms the vector $\mathbf{e}_f = \left[\begin{array}{ccc}
\mathbf{0}_{f-1} & 1 & \mathbf{0}_{m-f}
\end{array}\right]^T$. Then, the user sends the query vectors $\mathbf{u}$ to nodes $1$ and $2$ and $\mathbf{u}+\mathbf{e}_f$ to nodes $3$ and $4$. Each node responds by projecting its data onto the query vector it receives. Therefore, the responses of nodes $1, \dots, 4$ are given by $\mathbf{u}^TA$, $\mathbf{u}^TB$, $\mathbf{u}^TA+\mathbf{u}^TB+a_f+b_f$, $\mathbf{u}^TA+2\mathbf{u}^TB+a_f+2b_f$, respectively. From the responses of the nodes, the user will be able to obtain privately file $X_f = (a_f,b_f)$. We  measure the efficiency of a PIR scheme by its relative download cost referred to as {\em communication price of privacy $(cPoP)$}. To retrieve $2$ file symbols, the scheme downloads $4$ implying $cPoP=4/2=2$, which is asymptotically optimal (as $m\rightarrow\infty$) \cite{banawan2016capacity, chan2014private}.
\end{example}


In the previous scheme in Example~\ref{sec:introex} and its generalization in \cite{tajeddine2016private, extended}, the user needs to wait for  the responses of all the nodes to be able to decode the file. However,  this may not be possible in many cases due to some nodes being   unresponsive or due to network  failures. Even when all the nodes are responsive, some of them  may be slow (due to being busy or due to a slow connection). A single slow node will delay the user, even if all the other nodes are fast. In this case, it may be better for the user to ``cut" the slow node, consider it unresponsive and re-query the fast nodes. We are interested in constructing PIR schemes that have this {\em adaptive} property.

 A PIR scheme that can work even in the presence of unresponsive  servers has been studied in the literature in the case of replicated data, and is called robust PIR scheme \cite{augot2014storage, beimel2003robust, devet2012optimally}. We say that a  PIR scheme is  $\nu$-robust if it can tolerate $\nu$ unresponsive or slow servers. The standard method for achieving robustness is to design the queries such that the nodes' responses contain enough redundancy to tolerate $\nu$ erasures. In analogy with the existing work in the literature, we aim at designing $\nu$-robust PIR schemes that can operate on {\em coded}  and not only on replicated data. However, we require the additional property that the scheme is {\em universally optimal}, in the  sense of achieving the minimum $cPoP$ simultaneously for any number  of unresponsive nodes up to $\nu$ of them. Therefore, we avoid having to design the scheme for the worst-case scenario assuming the maximum number of unresponsive nodes. 

\begin{contexample} Consider again the same setting as before.
We want to design a universal $1$-robust PIR scheme. We propose an adaptive  scheme with two layers. The first layer is the same one described in the previous part of this example. If there is no unresponsive nodes the scheme stops after the first layer. The second layer depends on which node is unresponsive, or deemed slow, and is described in table~\ref{tab:exint}.

\begin{table}[t]
\vspace{5em}
\centering
\begin{tabular}{cccccc}
 Node 1 & $\mathbf{u}$\tikz[overlay,>=stealth]{\draw [decorate,decoration={brace}](-25pt,30pt) -- node[above,yshift=1mm, xshift = 1.1mm,font=\scriptsize]{Layer $1$} ++(37pt,0);}  &\tikz[overlay,>=stealth]{\draw [decorate,decoration={brace}](-15pt,30pt) -- node[above,yshift=1mm, xshift = 1.1mm,font=\scriptsize]{Layer $2$} ++(145pt,0);} $\varnothing$ \tikz[overlay,>=stealth]{\draw [decorate,decoration={brace}](-25pt,10pt) -- node[above,yshift=1mm, xshift = 0mm,font=\tiny]{\begin{varwidth}{0.95cm}{Node $1$ is unresponsive}\end{varwidth}} ++(35pt,0);} & $\mathbf{v}$\tikz[overlay,>=stealth]{\draw [decorate,decoration={brace}](-19pt,10pt) -- node[above,yshift=1mm, xshift = 0.5mm,font=\tiny]{\begin{varwidth}{0.95cm}{Node $2$ is unresponsive}\end{varwidth}} ++(33pt,0);} &$\mathbf{v}+\mathbf{e}_f$\tikz[overlay,>=stealth]{\draw [decorate,decoration={brace}](-30pt,10pt) -- node[above,yshift=1mm, xshift = 0.5mm,font=\tiny]{\begin{varwidth}{0.95cm}{Node $3$ is unresponsive}\end{varwidth}} ++(35pt,0);} & $\mathbf{v}+\mathbf{e}_f$\tikz[overlay,>=stealth]{\draw [decorate,decoration={brace}](-31pt,10pt) -- node[above,yshift=1mm, xshift = 0.5mm,font=\tiny]{\begin{varwidth}{0.95cm}{Node $4$ is unresponsive}\end{varwidth}} ++(33pt,0);}\\ 
  Node 2 & $\mathbf{u}$ & $\mathbf{v}$ & $\varnothing$ & $\mathbf{v}$ & $\mathbf{v}$ \\ 
  Node 3 & $\mathbf{u}+\mathbf{e}_f$ & $\mathbf{v}+\mathbf{u}$ & $\mathbf{v}+\mathbf{u}$ & $\varnothing$ & $\mathbf{v}$\\
  Node 4 & $\mathbf{u}+\mathbf{e}_f$ &  $\mathbf{v}$ & $\mathbf{v}$ & $\mathbf{v}$ & $\varnothing$
 \end{tabular}
 \captionof{table}{An example of our proposed $1$-universal and adaptive robust PIR scheme. The scheme has two layers, with $\varnothing$ indicating the unresponsive node.}
 \vspace{-0.75em}
 \label{tab:exint}
 \end{table}
 
Suppose, for example, that node $1$ is not responsive. In this case, the user will be missing  $\mathbf{u}^TA$ (the response of node $1$) and needs it to be able to decode using the $3$ other responses from the first layer. The goal of the second layer is to retrieve $\mathbf{u}^TA$ or another linear combination that  allows full decoding in the first layer. Only nodes $3$ and $4$ can give $\mathbf{u}^TA$, but if the user  asks directly for it in Layer $2$  it will reveal $\mathbf{e}_f$ to the node and therefore the identity of the requested file. That's why the user generates a new random vector $\mathbf{v}=\left[\begin{array}{ccc}
v_1 & \dots &v_m
\end{array}\right]^T$ with elements $\in GF(3)$. Implementing the queries in the second column in Table~\ref{tab:exint},  
the user can  decode $\mathbf{u}^TA+\mathbf{u}^TB$ in layer $2$, and then $X_f$ using the responses from  layer $1$. This schemes achieves asymptotically optimal $cPoP$ simultaneously for $0$ unresponsive nodes ($cPoP=2$) and $1$ unresponsive node ($cPoP=3$), as given in \eqref{eq:1} explained later. 


%

\end{contexample}

\noindent{\em{Related work:}} 
Until recently, most of the work on PIR has focused on replicated data and minimizing the total download cost \cite{beimel2001information, beimel2002breaking,dvir20142, yekhanin2008towards,yekhanin2010private, sun2016capacitynoncol, sun2016capacity, efremenko20123}. 
Recent work has studied PIR schemes on coded data. It was shown  in \cite{shah2014one} that downloading one extra bit is enough to achieve privacy, if the number of servers is exponential in the number of files. In \cite{chan2014private}, the authors derive bounds on the tradeoff between storage cost and download cost for linear coded data. Later, the authors in \cite{banawan2016capacity} derive the optimal lower bounds on download cost. Methods for transforming PIR schemes with replicated data to schemes on coded data were devised in \cite{fazeli2015pir}. This work was later generalized to PIR array codes in \cite{blackburn2016pir}. PIR schemes for MDS coded data 
were presented in \cite{tajeddine2016private, extended}. For the case of non-colluding nodes, these schemes achieve  asymptotically optimal download cost. A new family of PIR schemes on MDS coded data was constructed in \cite{freij2016private}, which achieves a lower download cost then the ones in \cite{tajeddine2016private} for the case of colluding nodes. In terms of fundamental limits, it was shown in \cite{sun2016capacitynoncol} that the so-called  PIR capacity  is $(1+1/n+1/n^2+\dots+1/n^{m-1})^{-1}$, which implies optimal $cPoP = 1+1/n+1/n^2+\dots+1/n^{m-1}$, where $n$ is the number of nodes and $m$ is the number of files. This capacity expression was then generalized to the case of a fixed number of colluding nodes in \cite{sun2016capacity}. All the previous fundamental results are for replicated data.  When the data is  coded using an $(n,k)$ MDS code, it was shown in \cite{banawan2016capacity} that the optimal $cPoP$ is $1+k/n+k^2/n^2+\dots+k^{m-1}/n^{m-1}$, thus the asymptotically optimal $cPoP=\frac{n}{n-k}$, as the number of files $m$ goes to infinity.
The setting in which nodes can be byzantine (malicious) and store replicated data was considered in \cite{augot2014storage, beimel2003robust, devet2012optimally} and robust PIR schemes were devised using locally decodable codes. 

\noindent{\em Contributions:}
In this paper, we present a construction of universal $\nu$-robust PIR schemes on $(n,k)$ MDS coded data, where $\nu$ is  the maximum number of unresponsive nodes\footnote {The parameter $\nu$ can be between $0$ and $n-k-1$. A $0$-robust scheme is a non-robust scheme.  If $\nu=n-k$, i.e., there is no redundant data queried, then perfect privacy can not be achieved except by downloading all the files.  If $\nu>n-k$, the file can not be fully retrieved since the MDS code cannot tolerate more than $n-k$ failures.}. We focus on non-colluding nodes (i.e., no spy nodes in the model in \cite{tajeddine2016private,extended}) and want to achieve perfect privacy which guarantees that zero information is leaked to the individual nodes about the index of the retrieved file. The construction is a generalization of our PIR schemes on MDS codes in \cite{tajeddine2016private}, with robustness against up to $\nu$ unresponsive nodes. The proposed scheme consists of two layers and has the following properties: (i) {\em universality}, meaning the scheme allows the user to retrieve the requested file privately, for all number of unresponsive servers up to $\nu$, and achieving the optimal $cPoP = \frac{n-i}{n-i-k}$ for all $i=1,\dots, \nu$, where $i$ is the actual number of unresponsive nodes; and (ii) {\em adaptivity}, meaning the scheme changes depending on which nodes do not respond.


%

\section{System Model and Main Result}

We adopt the same model  in \cite{tajeddine2016private} and summarize it here.

\noindent{\em DSS:}
We consider a distributed storage system (DSS) formed of $n$ nodes indexed from $1$ to $n$. The DSS stores $m$ files, $X_1,\dots,X_m$ using an $(n,k)$ MDS code over $GF(q)$, which achieves reliability against $n-k$ node failures.
Each file $X_i$ is divided into $k$ blocks, and each block is divided into $\alpha$ stripes or subdivisions. Thus, a file $X_i$ could be represented by a $k\times \alpha$ matrix with symbols chosen from the finite field $GF(q^\ell)$. 
The stripes are considered to be encoded separately using the generator matrix of the same MDS code. We assume the user knows the encoding vector used to encode the data on each node. We denote the column vector stored on node $i$ by $W_i\in GF(q^\ell)^{m\alpha}$. For instance, in Example~\ref{sec:introex}, $\alpha=1$, $W_1 = A, W_2 = B, W_3 = A+B, W_4 = A+2B,$ and $q=5.$ We assume the MDS code code is given and is not a design parameter.
  
%


\noindent{\em PIR:} 
The user wants to retrieve file $X_f$, from $n$ nodes, privately, meaning without revealing the index, $f$, to any of the nodes. We assume that the nodes in the DSS do not collude and that  $f$ is chosen uniformly at random from the set $\{1,\dots, m\}$.  We say that a PIR scheme over $GF(q)$ is linear, and of dimension $d$, when the request sent to node $i$ is a $d\times m\alpha$ query matrix, $Q_i$, over $GF(q)$. In this case, the response of  a node  is the projection of its data onto the query matrix. We want the PIR scheme to achieve perfect privacy, i.e., $H(f|Q_i)=H(f)$, for all $i$. Here, $H(.)$ denotes the entropy function.




\begin{definition}[Universal $\nu$-robust PIR scheme]
A universal $\nu$-robust PIR scheme is a PIR scheme which can tolerate up to $\nu$ unresponsive nodes, and for any  number of unresponsive nodes $0 \leq i \leq \nu$, it achieves perfect privacy with minimum $cPoP$ given by  (assuming no node collusion) \vspace{-0.5em} \begin{equation}
cPoP = \frac{n_i}{n_i-k}, \label{eq:1}
\end{equation}
where $n_i = n-i$ is the number of responsive nodes.
\end{definition}





Theorem~\ref{th:main} gives the   main result of this paper and is proved   in Section~\ref{sec:proof}.

\begin{theorem} \label{th:main}
Consider a DSS with $n$ non-colluding nodes and using an $(n,k)$ MDS code over $GF(q)$. Then, the linear PIR scheme over $GF(q)$ described in Section~\ref{sec:constructionb1} is a universal $\nu$-robust PIR scheme, i.e., it achieves perfect privacy and and has optimal  $cPoP=\frac{n_i}{n_i-k}$, where $n_i=n-i$, for all number of unresponsive nodes $i, 0\leq i\leq\nu$.
\end{theorem}

\section{Robust PIR  Scheme Description}\label{sec:constructionb1}

 We describe here  the universal  PIR scheme referred to in Theorem~\ref{th:main}. This scheme is adaptive and  consists of two layers.


{\bf A. Layer $1$} is essentially multiple copies of the non-robust PIR scheme of Theorem~1 in \cite{tajeddine2016private}. This scheme requires a number of subdivisions $\alpha = \frac{LCM(k,n-k)}{k}$ and is of dimension $d'=\frac{LCM(k,n-k)}{n-k}$, i.e., it consists of $d'$ subqueries.

WLOG, we assume the code is systematic and write $n-k = \beta k + r$, where $\beta$ and $r$ are integers and  $0\leq r <k$ and $\beta \geq 0$. We  divide the nodes into groups, as seen in Table~\ref{tab:q}.   The first group consists of $k$ nodes and is divided into two sub-groups.  The first consists of $r$ nodes, which are chosen to be the first $r$ nodes for the first subquery. The second is formed of the remaining $k-r$ nodes. As for the parity nodes, we divide them into $\beta$ groups of $k$ nodes each, and one group of $r$ nodes.

Table~\ref{tab:q} describes the first subquery of the PIR scheme when the user wants file $X_f$. The user generates a random vector $\mathbf{u}$, whose elements are chosen uniformly at random from $GF(q)$, the same field over which the MDS code is defined.
Next, we summarize how the remaining subqueries are constructed. For each subquery $j, j = 2, \dots, d'$, a new random vector $\mathbf{v}_j$ is created. The subqueries to the first group, assumed to be systematic, are shifted cyclically downwards in each subquery. As for the  remaining $\beta $ groups of $k$ nodes each, the query to each group $s$ is  $\mathbf{v}_j+\mathbf{e}_{(f-1)\alpha + r+(s - 2)d+j}$, where vector $\mathbf{e}_j$ is the all-zero vector with a single $1$ in position $j$. As for the last $r$ nodes, the random vector $\mathbf{v}_j$ is sent in subquery $j$.
\ADLinactivate\begin{table}
\centering
\begin{tabular}{|c|c|}
\hline
Nodes & Queries \\\hhline{|=|=|}
\tikz[overlay,>=stealth]{\draw [decorate,decoration={brace,mirror}](-29pt,6pt) -- node[left,rotate = 90,yshift=3mm, xshift = 4.5mm,font=\tiny]{Group $1$} ++(0,-80pt);}$1$ & $\mathbf{u}+\mathbf{e}_{(f-1)\alpha + 1}$ \\
$2$ & $\mathbf{u}+\mathbf{e}_{(f-1)\alpha + 2}$ \\
\vdots & \vdots \\
$r$ & $\mathbf{u}+\mathbf{e}_{(f-1)\alpha + r}$ \\\hline
$r+1$ & \multirow{4}{*}{$\mathbf{u}$} \\
\vdots & \\
$k$ & \\\hline
\tikz[overlay,>=stealth]{\draw [decorate,decoration={brace,mirror}](-21.5pt,6pt) -- node[left,rotate = 90,yshift=3mm, xshift = 5mm,font=\tiny]{Group $2$} ++(0,-35pt);}$k+1$ & \multirow{4}{*}{$\mathbf{u}+\mathbf{e}_{(f-1)\alpha + r+1}$} \\
\vdots & \\
$2k$ & \\\hline
\vdots & \vdots \\\hline
\tikz[overlay,>=stealth]{\draw [decorate,decoration={brace,mirror}](-19pt,6pt) -- node[left,rotate = 90,yshift=2.5mm, xshift = 6.5mm,font=\tiny]{Group $\beta+1$} ++(0,-35pt);}$\beta k+1$ & \multirow{4}{*}{$\mathbf{u}+\mathbf{e}_{(f-1)\alpha + r+(\beta-1)d+1}$} \\
\vdots & \\
$(\beta+1)k$ & \\\hline
\tikz[overlay,>=stealth]{\draw [decorate,decoration={brace,mirror}](-9pt,6pt) -- node[left,rotate = 90,yshift=2.5mm, xshift = 6.5mm,font=\tiny]{Group $\beta+2$} ++(0,-35pt);}$(\beta+1) k+1$ & \multirow{4}{*}{$\mathbf{u}$} \\
\vdots & \\
$n$ & \\\hline
\end{tabular}
  \caption{\label{tab:scheme} First subquery, in the non-robust PIR   scheme  in \cite{tajeddine2016private} (no collusion),   assuming the user wants file $X_f$.}
  \vspace{-0.5em}
  \label{tab:q}
\end{table}\ADLactivate

The number of subdivisions for a non-robust scheme on an $(n_i,k)$ MDS code is $\alpha_i = \frac{LCM(k,n_i-k)}{k},$ and the dimension or number of subqueries of the PIR scheme is $d'_i=\frac{LCM(k,n_i-k)}{n_i-k}.$

To achieve a universal $\nu$-robust PIR scheme on $(n,k)$ MDS code, we need enough ``granularity" to account for the different number $i$ of unresponsive nodes, for $i=0,\dots, \nu$.
The number of subdivisions $\alpha$ for a universal $\nu$-robust PIR scheme is the $LCM$ of the number of subdivisions, $\alpha_i$, of the scheme for all possible numbers of responsive servers $n_i$.

\vspace{-1em}
\begin{equation}
\alpha = LCM(\alpha_1,\dots,\alpha_i).\label{alph}
\end{equation}

\vspace{-0.5em}
The number of subqueries sent in layer $1$ is $d_0 = d'_0 \times \frac{\alpha}{\alpha_0}.$
For this, $d_0$ random vectors $\mathbf{u}_1, \dots, \mathbf{u}_{d_0}$ are created (one random vector per subquery). Every $\alpha_0$ subdivisions are queried in a set of $d'_0$ subqueries.

{\bf B. Layer $2$} depends on which nodes are unresponsive. Thus, the user will cut those nodes off and compensate for the responses from these nodes using extra subqueries to the other $n_i$ nodes. The goal of layer 2 is to allow the user to recover the responses that were missed in layer 1. However, this should be accomplished without violating the privacy constraint. Let us suppose $n_i$ is the number of responsive nodes in layer~$1$. Hence, in each copy of the scheme in layer $1$, there are $(n_0-n_i) \times d_0$ sub-responses missing. The goal of layer $2$ is to recover these sub-responses. We will divide the missing parts into $d_i-d_0$ groups of size $n_i-k$. Each of these groups will be asked for in one subquery, in the way a subquery is sent in an $(n_i,k)$ system to decode $n_i-k$ parts in \cite{tajeddine2016private}. 
There are two cases:
\begin{itemize}
\item Case 1: If the missing part is a function of $\mathbf{e}_j$ the user shall send $\mathbf{e}_j+\mathbf{u}_{s}$, where $\mathbf{u}_s$ is a new random vector, to one of the responsive nodes which never received a query on $\mathbf{e}_j$ in the previous subqueries.
For instance, if the required file is $X_1$, $\mathbf{e}_1, \dots, \mathbf{e}_r$ are asked for in group $1$, so to ask for them in the second layer, the user should ask them from any group other than group 1.
\item Case 2: On the other hand, if the missing part is a ``purely" randomvector ($\mathbf{u}_j$, for any $1\leq j\leq d_0$), the user shall send $\mathbf{u}_j+\mathbf{u}_s$ to one of the responsive nodes which never received a query $\mathbf{u}_j$ in the previous subqueries. 
For instance, to ask for a pure random vector $\mathbf{u}_i$ in layer $2$, it should not be queried from group $\beta$. Also, it should not be asked for from the $k-r$ nodes in group $1$ that have been asked for a purely random vector in subquery $j$.
\end{itemize}


After setting those, the random vector $\mathbf{u}_i$ will be sent to the rest of the nodes ($k$ nodes), in this subquery $i$.


\begin{table*}[t]
\vspace{1em}
\centering
  \begin{tabular}{|c|c|c|c|c|c|c|c|}
   \hline
node 1 & \tikz[overlay,>=stealth]{\draw [decorate,decoration={brace}](-5pt,10pt) -- node[above,yshift=1mm, xshift = 1.1mm,font=\scriptsize]{Layer $1$} ++(82pt,0);} $\mathbf{u}_1+\mathbf{e}_1$ & $\mathbf{u}_2$ & \tikz[overlay,>=stealth]{\draw [decorate,decoration={brace}](-16pt,10pt) -- node[above,yshift=1mm, xshift = 0.2mm,font=\scriptsize]{Layer $2$} ++(7.27cm,0);} $\varnothing$ &   $\mathbf{u}_3 + \mathbf{u}_1$ & $\mathbf{u}_3+\mathbf{e}_2$ & $\mathbf{u}_3+\mathbf{e}_2$ & $\mathbf{u}_3+\mathbf{u}_1$ \\\hline
    node 2 & $\mathbf{u}_1$ & $\mathbf{u}_2+\mathbf{e}_1$ & $\mathbf{u}_3 + \mathbf{u}_2$ & $\varnothing$ & $\mathbf{u}_3+\mathbf{e}_3$ & $\mathbf{u}_3+\mathbf{e}_3$ & $\mathbf{u}_3+\mathbf{u}_2$\\\hline 
    node 3 & $\mathbf{u}_1+\mathbf{e}_2$ & $\mathbf{u}_2+\mathbf{e}_3$ & $\mathbf{u}_3+\mathbf{e}_1$ & $\mathbf{u}_3+\mathbf{e}_1$ & $\varnothing$ &  $\mathbf{u}_3$ &  $\mathbf{u}_3$ \\\hline
    node 4 & $\mathbf{u}_1+\mathbf{e}_2$ & $\mathbf{u}_2+\mathbf{e}_3$ & $\mathbf{u}_3$ & $\mathbf{u}_3$ & $\mathbf{u}_3$ & $\varnothing$ &  $\mathbf{u}_3$ \\\hline
    node 5 & $\mathbf{u}_1$ & $\mathbf{u}_2$ & $\mathbf{u}_3$ & $\mathbf{u}_3$ & $\mathbf{u}_3$ &  $\mathbf{u}_3$ & $\varnothing$ \\\hline
  \end{tabular}
  \captionsetup{justification=centering}
  \caption{Queries to the nodes when one node is unresponsive. The 2 columns in Layer 1 represent the queries to all the nodes. Depending on which node is unresponsive (designated with $\varnothing$), one column in Layer 2 is chosen to query. }
  \label{tab:quer4}
\end{table*}

\subsection{Example on Scheme Construction}\label{sec:ex}

\begin{example}[Universal 2-robust PIR]\label{ex:2}
Consider the $(5,2)$ systematic MDS code storing $m$ files. Nodes $1,2,\dots 5$ store $A, B, A+B, A+2B, A+3B$, respectively, where $A$ and $B$ are as defined in example~\ref{sec:introex}. We want a universal $2$-robust PIR scheme, a PIR scheme that will be optimal in terms of communication price of privacy ($cPoP$) if $5$ nodes, $4$ nodes, and $3$ nodes respond. Let us call $n_0=5, n_1=4, n_2=3$.


We consider the number of subdivisions and dimension of each code. For the $(5,2)$ code, $\alpha_0 = 3$ and $d'_0 = 2$, for the $(4,2)$, $\alpha_1 = 1$ and $d'_1 = 1$, and for $(3,2)$, $\alpha_2=1$ and $d'_2 = 2$.

For the code to tolerate failures, we need to subdivide the files into $\alpha = LCM(\alpha_0,\alpha_1, \alpha_2) = 3.$

The number of subqueries required in order to retrieve the $\alpha$ parts of the file for code $(n_i,k)$ will be 

\vspace{-0.5em}
\begin{equation}
d_i=d'_i\times \frac{\alpha}{\alpha_i}. \label{di}
\end{equation}\vspace{-1em}

Thus, $d_0 = 2, d_1 = 3, d_2 =6.$

\noindent{\em Layer 1:}
Suppose the user wants $X_1$.
The user first sends subqueries to the $5$ nodes expecting  all of them to respond. We use here the  PIR scheme in \cite{tajeddine2016private} for the case of  a $(5,2)$ MDS code. The user creates random vectors $\mathbf{u}_1$, $\mathbf{u}_2$, and send the queries as in layer $1$ of table~\ref{tab:quer4}.


\noindent{\em Layer 2:} 

\noindent{\em Case 1:}
If one node does not respond, the user compensates for the missing information and sends an extra query to the other $4$ nodes. In one query to a $(4,2)$ system, the user can decode privately $2$ parts. In a $(4,2)$ system each query can give us 2 parts, thus $1$ extra query can compensate for the unresponsive node, this matches the number of subqueries being $3$. We generate a new random vector $\mathbf{u}_3$ for this extra query.

%
Column $i$ in layer 2 shows the subquery when node $i$ does not respond. For example, when node $1$ does not respond. We see that we are missing equation with $\mathbf{u}_1 + \mathbf{e}_1$, which is case $1$ and $\mathbf{u}_2$ which is case~$2$.
For this, we will send a $\mathbf{u}_3+\mathbf{u}_2$ to node $2$ and $\mathbf{u}_3 +\mathbf{e}_1$ to node 3. Of course, then we send $\mathbf{u}_3$ to nodes $1$ and $4$ to decode the interference. Table~\ref{tab:quer4} shows the sent queries.

\noindent{\em Case 2:}
On the other hand, if two nodes do not respond, then we need $4$ extra subqueries. For example, if nodes $1$ and $3$ do not respond, there are $4$ missing responses. Three of those missing parts are of case $1$, $\mathbf{e}_1, \mathbf{e}_2,$ and $\mathbf{e}_3$, and one missing part is of case $2$, $\mathbf{u}_2$. The queries in this case are shown in table~\ref{tab:quer3}.


\begin{table}[t]
\centering
\vspace{2em}
  \begin{tabular}{|c|c|c|c|c|c|c|}
   \hline
1 & \tikz[overlay,>=stealth]{\draw [decorate,decoration={brace}](-7pt,10pt) -- node[above,yshift=1mm, xshift = 1.1mm,font=\scriptsize]{Layer $1$} ++(85pt,0);} $\mathbf{u}_1+\mathbf{e}_1$ & $\mathbf{u}_2$ & \tikz[overlay,>=stealth]{\draw [decorate,decoration={brace}](-16pt,10pt) -- node[above,yshift=1mm, xshift = 1mm,font=\scriptsize]{Layer $2$} ++(5.78cm,0);} $\varnothing$ & $\varnothing$ & $\varnothing$ & $\varnothing$ \\\hline
2 & $\mathbf{u}_1$ & $\mathbf{u}_2+\mathbf{e}_1$ & $\mathbf{u}_3$ & $\mathbf{u}_4$ & $\mathbf{u}_5+\mathbf{e}_2$ & $\mathbf{u}_6+\mathbf{e}_3$ \\\hline 
3 & $\mathbf{u}_1+\mathbf{e}_2$ & $\mathbf{u}_2+\mathbf{e}_3$ & $\varnothing$ & $\varnothing$ & $\varnothing$ & $\varnothing$ \\\hline
4 & $\mathbf{u}_1+\mathbf{e}_2$ & $\mathbf{u}_2+\mathbf{e}_3$ & $\mathbf{u}_3 + \mathbf{e}_1$ & $\mathbf{u}_4 + \mathbf{u}_2$ & $\mathbf{u}_5$ & $\mathbf{u}_6$ \\\hline
5 & $\mathbf{u}_1$ & $\mathbf{u}_2$ & $\mathbf{u}_3$ & $\mathbf{u}_4$ & $\mathbf{u}_5$ & $\mathbf{u}_6$ \\\hline
  \end{tabular}
  \caption{Queries to the nodes when nodes $1$ and $3$ are unresponsive}
  \vspace{-1.5em}
  \label{tab:quer3}
\end{table}


\end{example}

\section{Proof of Theorem~\ref{th:main}}\label{sec:proof}

Before giving the proof of the theorem, we will state two properties of the PIR scheme in Theorem~1 in \cite{tajeddine2016private}  that will be essential to prove Theorem~\ref{th:main}.

\begin{property} \label{prop1}
In the PIR scheme in \cite{tajeddine2016private}, the query vectors in each sub-query can be permuted among the nodes without affecting the decodability and the privacy properties of the scheme. This follows directly from the fact that the node groups (see for e.g. Table~\ref{tab:scheme}) can be chosen arbitrarily.
\end{property}

\begin{property} \label{prop2}
The scheme in \cite{tajeddine2016private} allows the user to retrieve $\mathbf{e}_fX$, which is, in other words, the file $X_f$. This can be readily generlized to retrieve any $\mathbf{u}X$, where $\mathbf{u}$ is any vector of dimension $m$, where $m$ is the number of files.
\end{property}

Let us start by proving the decodability of the example in section~\ref{sec:ex}.

\subsection{Example~\ref{ex:2} decodability:}

Let $A$ and $B$ be as defined in example~\ref{sec:introex}.

\noindent{\em Layer $1$:}
The scheme's decodability when all nodes respond follows directly from \cite{tajeddine2016private}.

\noindent{\em Layer $2$:} We will prove that the scheme applied in section~\ref{sec:ex} is decodable when node 1, for example, is unresponsive. The nodes project the query vectors on the data they hold and send the response back to the user.

\begin{itemize}
\item We notice that from the third subquery, the user decodes the interference from nodes $3$ and $4$. Then gets $a_{11}+b_{11}$ and $\mathbf{u}_2^TB$. 
\item From the first subquery, $a_{12}+2b_{12}$ and $a_{12}+3b_{12}$ can be retrieved and thus decoding $a_{12}$ and $b_{12}$.
\item From $\mathbf{u}_2^TB$ and the response of node $4$ in the second query, the user decodes the interference $\mathbf{u}_2^TA$ and $\mathbf{u}_2^TB$. The user can then retrieve $b_{11}, a_{13}+b_{13}$, and $a_{13}+2b_{13}$. 
From these equations, along with $a_{11}+b_{11}$ retrieved from the third subquery (second layer), the user decodes $a_{11}, b_{11}, a_{13},$ and $b_{13}$.
\end{itemize}

Thus decoding all parts of the file $1$.
The $cPoP$ of the scheme if one node is unresponsive is $12\times\frac{1}{6} = 2$ which is the same as the optimal $cPoP$ found in \cite{banawan2016capacity}.
When $2$ nodes do not respond, if we look at the query table~\ref{tab:quer3}, we notice that the missing parts are retrieved, and achieve $cPoP = 3$.

\subsection{Decodability:}

\noindent{\em Layer $1$:} If $n_0 = n$ nodes respond, the decodability follows from \cite{tajeddine2016private}. Every $\alpha_0$ parts are decoded in $d'_0$ subqueries, thus retrieving the complete file (i.e. $\alpha$ parts) in $d_0 = \frac{d'_0\alpha}{\alpha_0}$ subqueries.

\noindent{\em Layer $2$:} If $n_i$ out of the $n$ nodes respond, $s_{i} = (n_0 - n_i) \times d_0$ responses are missing. In each extra subquery, the user can decode $n_i-k$ parts. Thus in total, from the $d_i-d_0$ extra subqueries the user can decode $(n_i-k)\times(d_i-d_0)$ parts. We can see that
$(n_i-k)(d_i-d_0) = d_0\times(n_0-n_i) = s_{i},$
by substituting the value of $d_i$ by its expression in~\eqref{di}. This shows that the number of decodable parts from the extra $d_i-d_0$ subqueries is equal to the number of missing parts.

Now the question is whether  the new sub-responses are able to provide parts that are sufficient for the user to be able to retrieve the file he/she wants.

Here, we use properties~\ref{prop1} and \ref{prop2}. Using property~\ref{prop1}, we can see that we can, in fact, decode $n_i-k$ parts in each subquery of this layer, since those subqueries are similar to the schemes in \cite{tajeddine2016private}, only permuted. Using property~\ref{prop2}, we can see that if a missing response is a function of a random vector in a subquery of layer $1$, the user can hide the random vector using the scheme in \cite{tajeddine2016private} and retrieve a new function, in layer $2$, that could substitute the missing response.

When all nodes are responsive, the responses from layer $1$ form a set of $k$ independent equations about the $k$ blocks of each stripe and about the interference, allowing the user to decode the file. However, when some nodes are unresponsive, some sub-responses, and therefore equations, are not retrieved. In this layer, the extra subqueries should be able to provide equations that will substitute those missing equations. Thus, those extra sub-responses, along with the sub-responses from layer $1$, should form a system of $k$ independent equations about each stripe and interference. 

The number of unresponsive nodes can be at most $\nu = n-k-1$. The number of independent equations required to retrieve a full stripe or to decode the interference is $k$. To form a solvable system of equations about a stripe, a node is asked at most once about the same stripe. Thus, the number of missing sub-responses about a certain stripe or interference is at most $\min(k, n-k-1)$. Consider the number of unretrieved equations about a stripe is $\gamma$. Since there are at least $k+1$ responding servers, there will always be at least $\gamma$ nodes that have not been asked for equations about this stripe before. The user can query those nodes to retrieve new equations about this stripe to substitute the missing ones.

\noindent{\em Privacy:} In each subquery, the query to a node is either one-time padded by an independent vector or the independent vector itself. Therefore,  the privacy of the scheme follows from the fact that the nodes do not collude.

\noindent{\em Optimality:} The price of privacy is optimal for any number of responsive nodes $n_i\geq n_{\nu}$, $cPoP = \frac{d_i n_i}{k\alpha}= \frac{n_i}{n_i-k}$, obtained by substituting equations \eqref{alph} and \eqref{di} into this equation.


\section{Conclusion}
We studied the problem of constructing robust PIR schemes with low communication cost for requesting data from a DSS storing data using MDS codes. The responses from certain nodes may be very slow. In such case, the user would cut those nodes off and ask the rest of the nodes for what he/she should have received from this node. The objective is to allow the user to do this with low communication cost.
We constructed adaptive universal $\nu$-robust PIR schemes with non-colluding nodes achieving the optimal price of privacy for all numbers of responsive nodes. The next steps would be to look into non-adaptive schemes, and schemes for colluding nodes.

\bibliographystyle{ieeetr}
\bibliography{coding2,coding1,}

\end{document}